\documentclass[12pt]{article}
\usepackage{amsmath,,calrsfs}
\usepackage{amsfonts}
\usepackage{amssymb}
\usepackage{amscd}
\usepackage{bbm}
\usepackage{fancybox}
\usepackage{cite}
\usepackage{amsmath,amsfonts,amsbsy}
\usepackage{pstricks,pst-node}
\usepackage[small,bf,hang]{caption2}
\usepackage{graphicx}
\usepackage{epsfig}
\usepackage{psfrag}
\usepackage{comment}

\usepackage{float}

\psset{unit=1.3cm,linewidth=.5pt,radius=.2}  

\usepackage{multirow}                     
\usepackage{float}                          
\usepackage{lscape}                         
\usepackage{bm}

\newcommand{\bb}{\bar\beta}

\newcommand{\beq}{\begin{equation}}
\newcommand{\eeq}{\end{equation}}
\newcommand{\bi}{\begin{itemize}}
\newcommand{\ei}{\end{itemize}}
\newcommand{\bt}{\begin{tabular}}
\newcommand{\et}{\end{tabular}}
\newcommand{\bc}{\begin{center}}
\newcommand{\ec}{\end{center}}

\newcommand{\be}{\begin{equation}}
\newcommand{\ee}{\end{equation}}
\newcommand{\bea}{\begin{eqnarray}}
\newcommand{\eea}{\end{eqnarray}}
\newcommand{\ba}{\begin{array}}
\newcommand{\ea}{\end{array}}

\def\bbox{{\,\lower0.9pt\vbox{\hrule \hbox{\vrule height 0.2 cm
\hskip 0.2 cm \vrule height 0.2 cm}\hrule}\,}}
\newcommand{\dsl}{\pa \kern-0.5em /}

\font\mybb=msbm10 at 12pt
\def\bb#1{\hbox{\mybb#1}}

\def\bQ {\bb{Q}}
\def\bJ {\bb{J}}

\def\bJ {\bb{J}}

\def\bfz{\mbox{\boldmath $\zeta$}}

\makeatletter \@addtoreset{equation}{section} \makeatother

\def\slashchar#1{\setbox0=\hbox{$#1$}           
   \dimen0=\wd0                                 
   \setbox1=\hbox{/} \dimen1=\wd1               
   \ifdim\dimen0>\dimen1                        
      \rlap{\hbox to \dimen0{\hfil/\hfil}}      
      #1                                        
   \else                                        
      \rlap{\hbox to \dimen1{\hfil$#1$\hfil}}   
      /                                         
   \fi}

\begin{document}

\begin{titlepage}
\begin{center}

\hfill  DAMTP-2016-38

\vskip 1.5cm

{\Large \bf Noether theorems and higher derivatives}

\vskip 1cm

{\bf Paul K. Townsend} \\

\vskip 25pt

{\em  \hskip -.1truecm
\em  Department of Applied Mathematics and Theoretical Physics,\\ Centre for Mathematical Sciences, University of Cambridge,\\
Wilberforce Road, Cambridge, CB3 0WA, U.K.\vskip 5pt }

{email: {\tt p.k.townsend@damtp.cam.ac.uk}} \\

\end{center}

\vskip 0.5cm

\begin{center} {\bf ABSTRACT}\\[3ex]
\end{center}

A simple proof of Noether's first theorem involves the promotion of a constant symmetry parameter $\epsilon$  to an arbitrary function of time; 
the Noether charge $Q$ is then the coefficient of $\dot\epsilon$ in the variation of the action.  Here we examine 
the validity of this proof for  Lagrangian mechanics with arbitrarily-high time derivatives, in which context ``higher-level'' analogs of  Noether's theorem can be
similarly proved, and ``Noetherian charges'' read off from, e.g. the coefficient of $\ddot \epsilon$ in the variation of the action. 
While $Q=0$ implies  a restricted gauge invariance,  unrestricted gauge  invariance requires zero Noetherian  charges too. Some illustrative examples  
are considered and the extension to field theory discussed.

\end{titlepage}
\section{Introduction}

Noether's theorem (or ``first theorem'') relates symmetries of an action to conservation laws \cite{Noether:1918zz}. It is easily proved for ``internal'' symmetries but the standard  proof for  spacetime  symmetries is more involved.  A simpler alternative proof  starts from the observation (in the context of mechanics) that if an action $I$ is invariant under an infinitesimal  transformation with constant  parameter $\epsilon$ then its variation for  {\it non-constant} $\epsilon$ is $\delta_\epsilon I = \int\! \dot \epsilon  Q dt$ for some  $Q$.  For suitable boundary conditions on $\epsilon$,  the left hand side is zero for solutions of the equations of motion while the right hand side can be rewritten  as $-\int \! \epsilon \dot Q dt$, which is zero  for all $\epsilon(t)$ iff $\dot Q=0$. So  $Q$ is the ``Noether charge'', which may be read off from the coefficient of $\dot\epsilon$ appearing in the  calculation of $\delta_\epsilon I$ for arbitrary $\epsilon(t)$ \cite{GellMann:1960np}.  The analogous observation in the context of field theory (with Noether current replacing Noether charge) was extensively used in the early construction of  supersymmetric field theories  by means of an  order-by-order ``Noether procedure'', e.g. \cite{deWit:1975veh}, and the simplified proof itself has  now displaced the traditional one in some of the more recent texts on Quantum Field Theory; e.g. \cite{Peskin:1995ev,Weinberg:1995mt}.  

But is the simplified proof correct?  The above statement of it  for mechanics  is cavalier with boundary terms. The integration  by parts to convert  $\int\! \dot \epsilon Q dt$ to $-\int \! \epsilon \dot Q dt$ will 
produce a boundary term; this is absent if $\epsilon$ is zero at the integration  limits, but no non-zero constant $\epsilon$ satisfies this condition. Boundary terms will 
also appear in $\delta_\epsilon I$, and these will contain derivatives of  $\epsilon$ if $I$ is the action for a ``higher-derivative'' theory.   Noether's original proof  allowed for higher derivatives; is the simpler alternative proof equally general? And does it also apply when only the equations of motion are invariant, and not the action? 

These considerations led the author to an examination, in the context  of Lagrangian mechanics with a Lagrangian involving arbitrarily-high time derivatives,  of the precise relationship  between the traditional proof of Noether's  theorem and the simplified proof.  The principal conclusion is that the simplified proof goes through even  in the higher-derivative case  but the presence of higher derivatives makes it necessary to impose boundary conditions on some derivatives of $\epsilon$. 

Relaxation of these derivative boundary conditions on $\epsilon$ leads to the concept of ``higher-level'' Noether theorems for higher-derivative actions that are invariant under some transformation with parameter  $\epsilon$ not only when $\epsilon$ is constant but also when $\dot\epsilon$ is constant, and then when $\ddot\epsilon$ is constant too, etc.  In the simplest case of a Lagrangian involving  second (but no higher)  time derivatives, the variation of the action takes the form $\delta_\epsilon I = \int \ddot \epsilon\, \bQ_1dt$ for such symmetry transformations, and a straightforward extension of the simplified proof of Noether's theorem now yields the conclusion that $\ddot\bQ_1=0$ as a consequence of the equations of motion. Integration of this equation for the ``level-$1$ Noetherian charge'' $\bQ_1$ yields  two standard (level-zero) Noether charges: corresponding to the two symmetries  implicit in a symmetry parameter  $\epsilon(t)$ that is linear in $t$.  There is a straightforward extension to level-$p$ Noetherian charges $\bQ_p$ for Lagrangians that contain at least $(p+1)$th derivatives of the dependent variables.  This  idea is illustrated with a simple example. 

Given a symmetry transformation, with constant parameter $\epsilon$,  of an action {\it without higher derivative terms}, the vanishing of the corresponding Noether charge $Q$ implies  that the action is actually invariant for {\it arbitrary} $\epsilon(t)$; i.e. the  supposed symmetry is really a gauge invariance. In this case the equations of motion are subject to an identity relating them; this is Noether's ``second theorem''. 
This $Q=0$ condition for gauge invariance is well known but, as we show here,  it applies to higher-derivative actions only if the gauge transformation parameter $\epsilon(t)$ 
is restricted by boundary conditions on its derivatives. If one requires  {\it unrestricted} gauge invariance in this higher-derivative context  then both $Q$ {\it and} all higher-level Noetherian 
charges $\bQ_p$ must vanish; a simple level-1 example of this is given. 

A discussion of some further obvious questions raised by these results is relegated to a final section. One question concerns the field theory analog of  ``Noetherian charges'': Another  question concerns the possibility of  avoiding higher derivatives by the inclusion of additional auxiliary variables.

\section{Noether's theorem}
\label{sec:one}

We will consider a Lagrangian depending on a single real function $x(t)$,  defined for real (time) variable $t$  in an interval that contains the interval $(\alpha,\beta)$, and on the derivatives of $x(t)$ up to the $n$th order.
The action therefore takes the form 
\begin{equation}
I[x] = \int_\alpha^\beta \! L\left(x,\dot x,\ddot x, \cdots, \frac{d^n x}{dt^n}; t\right)\, dt\, . 
\end{equation}
The extension to an action functional of many functions is straightforward, and  
such an extension will be considered in section \ref{sec:gaugeinvariance} for some special cases. 

A general transformation of both the independent variable $t$ and the dependent variable $x$ takes the form
\begin{equation}\label{transform}
t\to t^* (t) \, , \qquad x(t) \to x^*(t^*)\, , 
\end{equation}
for some monotonic function $t^*$ (we assume  $\dot t^*>0$ since this  is true for  the infinitesimal form of all continuous transformations)
and some new dependent function $x^*$. As a result, 
\begin{equation}
I[x]\to I ^*[x^*] = \int_{\alpha^*}^{\beta^*}  L\left(x^*(t^*), \frac{dx^*(t^*)}{dt^*}, \cdots ,  \frac{d^n x^*(t^*)}{dt^*{}^n}; t^*\right)\, dt^* \, , 
\end{equation}
where
\begin{equation}
\alpha^*= t^*(\alpha)\, , \qquad  \beta^*= t^*(\beta)\, .
 \end{equation}
The transformation (\ref{transform}) is a symmetry of the action if $I^*[x^*] =I[x]$ for any choice of the interval $(\alpha,\beta)$.   Here we are following the  initial discussion of Noether's theorem as presented in  Chapter 4
of the book by Gelfand and Fomin \cite{GF}, except that we allow for a ``higher-derivative''  ($n>1$) Lagrangian.

For a continuous symmetry with a real parameter $\epsilon$ we may consider an infinitesimal transformation; i.e. $\epsilon$ is small and we consider only 
variations that are first-order small. In this case we have 
\begin{equation}\label{inf}
t^* =  t + \epsilon\xi(t)\, , \qquad x^*(t^*) = x(t) + \epsilon\zeta(t)\, , 
\end{equation}
for two functions $(\xi,\zeta)$. It then follows that 
\begin{equation}\label{active}
x^*(t) = x(t) + \epsilon h(t)\, , \qquad h= \zeta-\xi\dot x\, ,
\end{equation}
and
\begin{equation}\label{starlims}
\alpha^* = \alpha + \epsilon\xi(\alpha)\, , \qquad \beta^* = \beta + \epsilon\xi(\beta)\, .
\end{equation}

Returning to the integral for $I^*[x^*]$, we  now {\it relabel} the integration variable $t^*$ as $t$; this has no effect on the integration limits, so
\begin{eqnarray}
I^*[x^*] &=& \int_{\alpha^*}^{\beta^*} L\left(x^*, \dot x^*, \cdots , \frac{d^n x^*(t)}{dt^n}; t\right) dt  \nonumber \\
&=& \int_\alpha^\beta  L\left(x + \epsilon h, \dot x + \frac{d(\epsilon h)}{dt} , \cdots , x^{(n)} + \frac{d^n(\epsilon h)}{dt^n}; t\right) dt\,  + \left[\epsilon \xi L\right]_\alpha^\beta\, , 
\end{eqnarray}
where we have used  (\ref{active}) and (\ref{starlims})  to get to  the second line. Now we expand the integrand to first order in $\epsilon$ to get
\begin{equation}
I^*[x^*] - I[x] =  \sum_{k=0}^n  \int_\alpha^\beta \! \left\{\frac{d^k(\epsilon h)}{dt^k} \frac{\partial L}{\partial x^{(k)}}\right\}dt  \ +\  \left[\epsilon \xi L\right]_\alpha^\beta \, .
\end{equation}
Let us call the left hand side $\delta_\epsilon I$. Integrating by parts on the right hand side yields the result
\begin{equation}\label{DelI}
\delta_\epsilon I =   \int_\alpha^\beta\! \epsilon h \, \frac{\delta I[x]}{\delta x(t)} dt  \ + \ \left[\epsilon\xi L - \sum_{k=1}^n\sum_{j=1}^k (-1)^j \frac{d^{k-j}(\epsilon h)} {dt^{k-j}} \, \frac{d^{j-1}}{dt^{j-1}}\left(\frac{\partial L}{\partial x^{(k)}}\right)\right]_\alpha^\beta \, , 
\end{equation}
where 
\begin{equation}
\frac{\delta I}{\delta x} = \sum_{k=0}^n (-1)^k \frac{ d^k}{dt^k} \left(\frac{\partial L}{\partial x^{(k)}}\right)\, . 
\end{equation} 
The functional derivative is defined for appropriate boundary conditions on $x$, and then the Euler-Lagrange (EL) equation is $\delta I/\delta x=0$. 

So far we did not use the fact that $\epsilon$  is a constant.  However, if we now suppose that $\epsilon$ {\it is}  constant then we may rewrite (\ref{DelI}) as 
\begin{equation}\label{Qzero}
\delta_\epsilon I = \epsilon\left\{  \int_\alpha^\beta h \, \frac{\delta I[x]}{\delta x(t)} dt  +  \left[ Q_0\right]_\alpha^\beta\right\}\, , 
\end{equation}
where 
\begin{equation}\label{Qzero2}
Q_0 = \xi L - \sum_{k=1}^n\sum_{j=1}^k (-1)^j \frac{d^{k-j}h} {dt^{k-j}} \, \frac{d^{j-1}}{dt^{j-1}}\left(\frac{\partial L}{\partial x^{(k)}}\right)\, . 
\end{equation}
The left hand side of (\ref{Qzero})  is zero for a symmetry of the action, by definition, but we can allow for symmetries of the equations of motion that are {\it not}  symmetries of the action by supposing that 
\begin{equation}
\delta_\epsilon I =  \left[ \epsilon \Delta\right]_\alpha^\beta
\end{equation}
for some quantity $\Delta$, which will be linear in the functions $(\xi,\zeta)$. In this case we have
\begin{equation}\label{basic2}
0= \int_\alpha^\beta \left\{ h \, \frac{\delta I[x]}{\delta x(t)} dt  + \dot Q\right\}dt\, , 
\end{equation}
where 
\begin{equation}\label{Q}
Q=  \xi L  -  \sum_{k=1}^n\sum_{j=1}^k (-1)^j \frac{d^{k-j}h} {dt^{k-j}} \, \frac{d^{j-1}}{dt^{j-1}}\left(\frac{\partial L}{\partial x^{(k)}}\right) -  \Delta\, . 
\end{equation}
Since the interval $(\alpha,\beta)$ is arbitrary, it follows that  the functions $(\xi,\zeta)$ define a symmetry transformation iff\footnote{Strictly speaking, this is the condition for a symmetry only if $Q\ne0$ because we have a gauge invariance if $Q=0$, as will be discussed in section  \ref{sec:gaugeinvariance}.}
\begin{equation}\label{symcon}
 h \, \frac{\delta I[x]}{\delta x(t)} dt  + \dot Q \equiv 0\, .  
 \end{equation}
 For solutions of the EL equation, this reduces to $\dot Q=0$; in other words, $Q$ (the Noether charge) is a constant of the motion. This is the standard proof  of Noether's theorem except that we proceeded without the assumption that the parameter $\epsilon$ is constant until the final steps. 

As a check on the above formulae,  consider the following action,  which can be interpreted as a Wess-Zumino action for the Galilean group in one dimension \cite{deAzcarraga:1990gs}:
\begin{equation}
I[x]= \int \! \tfrac{1}{2} \dot x^2 dt \, . 
\end{equation}
The  $3$-parameter Galilean transformation,  for constants $(\xi_0, \zeta_0)$ and $v$,  is 
\begin{equation}
\xi(t) = \xi_0 \, , \quad \zeta(t) = \zeta_0 + vt \qquad \Rightarrow \quad \Delta = \dot\zeta x \equiv v x\, .  
\end{equation}
The formula (\ref{Q}) gives
\begin{equation}
Q=  \xi  \left(\tfrac{1}{2}\dot x^2\right)  + (\zeta - \xi\dot x) \dot x - \Delta = -\xi_0\left( \tfrac{1}{2} \dot x^2\right) + \zeta_0  \left(\dot x\right) + v\left(t\dot x -x\right)\, , 
\end{equation}
and a calculation  shows that (\ref{symcon}) is satisfied:
\begin{equation}
h \, \frac{\delta I[x]}{\delta x(t)}   + \dot Q  =   - \dot\xi \left(\tfrac{1}{2} \dot x^2\right) + \dot\zeta \dot x - \dot\Delta = 0\, . 
\end{equation}
It follows that  $Q$ is the the $3$-parameter Noether charge for the Galilean symmetry.

\subsection{Reading off the Noether charge}

Let us now return to (\ref{DelI}), which is valid for arbitrary $\epsilon(t)$,  and isolate the particular  boundary term that involves no derivatives of $\epsilon$. We find that 
\begin{equation}\label{needlater}
\delta_\epsilon I  =   \int_\alpha^\beta\!  \epsilon h \, \frac{\delta I[x]}{\delta x(t)} dt \ + \ \left[\epsilon Q_0\right]_\alpha^\beta + \sum_{\ell=1}^{n-1}\left[ \frac{d^\ell \epsilon}{dt^\ell} \bQ_\ell \right]_\alpha^\beta\, , 
\end{equation}
where
\begin{equation}\label{Qell}
\bQ_\ell = -\frac{1}{\ell !} \sum_{k=1}^n \sum_{j=1}^k (-1)^j \frac{(k-j)!}{(k-j-\ell)!} \left(\frac{d^{k-j-\ell}h}{dt^{k-j-\ell}}\right) \frac{d^{j-1}}{dt^{j-1}} \left(\frac{\partial L}{\partial x^{(k)}}\right)\, . 
\end{equation}
This is equivalent to 
\begin{equation}\label{needlater1}
\delta_\epsilon I  -\left[ \epsilon \Delta\right]_\alpha^\beta =   \int_\alpha^\beta\!  \epsilon h \, \frac{\delta I[x]}{\delta x(t)} dt \ + \ \left[\epsilon Q\right]_\alpha^\beta + \sum_{\ell=1}^{n-1}\left[ \frac{d^\ell \epsilon}{dt^\ell} \bQ_\ell \right]_\alpha^\beta\, , 
\end{equation}
which we can rewrite as 
\begin{equation}\label{rewrite}
\delta_\epsilon I  - \left[ \epsilon \Delta\right]_\alpha^\beta = \int_\alpha^\beta \left\{ \epsilon \left( h \, \frac{\delta I[x]}{\delta x(t)} + \dot Q\right) + \dot\epsilon Q\right\} dt + \sum_{\ell=1}^{n-1}\left[ \frac{d^\ell \epsilon}{dt^\ell} \bQ_\ell \right]_\alpha^\beta\, . 
\end{equation}
If the left hand side is zero for {\it constant} $\epsilon$  then (\ref{symcon}) holds and we have, for arbitrary $\epsilon(t)$,  
\begin{equation}\label{start}
\delta_\epsilon I  -\left[ \epsilon \Delta\right]_\alpha^\beta  = \int_\alpha^\beta\! \dot\epsilon Q \, dt +  \sum_{\ell=1}^{n-1}\left[ \frac{d^\ell \epsilon}{dt^\ell} \bQ_\ell \right]_\alpha^\beta\, . 
\end{equation}
The  boundary term  on the right hand side of this equation is  present only for $n>1$, i.e. for a higher-derivative theory. It might appear that this could create an ambiguity in the coefficient of 
$\dot\epsilon$ in the integral. For example, the boundary term $[\dot\epsilon \bQ_1]$ could be replaced by an addition of  $\dot\epsilon \dot \bQ_1 + \ddot\epsilon \bQ_1$ 
to the integrand of the integral (as we shall do in the following section). This changes the coefficient of $\dot\epsilon$ in the integral  but it  also introduces a term involving $\ddot\epsilon$. 
If  we insist on  an integrand of the form $\dot\epsilon Q$, which we can do by integration by parts where necessary, then the coefficient $Q$ of $\dot\epsilon$ in the integral is unique.
And, as we have already shown, this coefficient is the Noether charge. 

This result (that the Noether charge can be read off from the coefficient of $\dot\epsilon$)  is usually deduced from the simplified proof of Noether's theorem, which we take up below, but we now see that {\it it does not depend on the validity of that proof}.
This fact is potentially significant for higher-derivative theories because, as we shall see, the simplified proof then requires conditions on $\epsilon$ that we did not yet need to impose.

\subsection{The simplified proof}
\label{subsec:simplified}

We have  now confirmed the main practical implication of the simplified proof of Noether's theorem, but  what about the proof itself?  

Given a continuous symmetry of the action, with infinitesimal transformation specified  by the functions $(\xi,\zeta)$, the variation of the action after promotion of the  parameter $\epsilon$
to an arbitrary function of time is given by (\ref{start}). If we restrict the function $\epsilon(t)$ by the boundary conditions 
\begin{equation}\label{highd}
0= \left. \frac{d^\ell \epsilon}{dt^\ell}\right|_{t=\alpha,\beta}\, \qquad  \ell =1, \dots, n-1\, , 
\end{equation}
then (\ref{start}) reduces  to 
\begin{equation}\label{starting}
\delta_\epsilon I    = \int_\alpha^\beta\!  \dot\epsilon Q\,  dt + \left[ \epsilon \Delta\right]_\alpha^\beta \, .  
\end{equation}
This is  the form of $\delta_\epsilon I$ assumed in  the simplified proof of Noether's theorem as outlined in the first paragraph of the Introduction,  but generalised to allow for $\Delta\ne0$.   Notice that 
(\ref{highd}) actually restricts $\epsilon$ only when the action has higher-derivative terms; in their absence no
restriction on $\epsilon(t)$ is needed for the validity of (\ref{starting}). 

Now we integrate by parts and use $Q+\Delta=Q_0$ to arrive at
\begin{equation}\label{nearly}
\delta_\epsilon I   = - \int_\alpha^\beta\! \epsilon\dot  Q \, dt \, + \, \left[\epsilon Q_0\right]_\alpha^\beta\, . 
\end{equation}
It  can be seen from (\ref{needlater})  that $\delta_\epsilon I= [\epsilon Q_0]$ for a solution of the EL equations when the conditions (\ref{highd}) are satisfied, 
and it follows from this  that $\dot Q=0$ as a consequence of the EL equation.  However, this does {\it not}  qualify as a ``simplified''  proof of Noether's theorem because 
the steps leading to (\ref{needlater}) constitute  most of the work of the standard proof! 

The key simplification of the simplified proof is that we can avoid the need for a precise knowledge of the boundary term in $\delta_\epsilon I$  by 
choosing
\begin{equation}\label{bcbasic}
\epsilon(\alpha)=\epsilon(\beta)=0\, . 
\end{equation}
A concern raised in the Introduction was that  this excludes a non-zero constant $\epsilon$.  However,  (\ref{nearly})  is valid (given a solution of the EL equation) for {\it any} $\epsilon(t)$ 
(satisfying (\ref{highd}) when $n>1$) so we are free to follow the consequences of choosing $\epsilon(t)$ to  satisfy (\ref{bcbasic}). One consequence is that  (\ref{nearly}) reduces to 
\begin{equation}
\delta_\epsilon I  = - \int_\alpha^\beta\! \epsilon\dot  Q \, dt \, . 
\end{equation}
Another consequence is that  $\delta_\epsilon  I$ is zero for solutions of the EL equation, so the EL equation implies that $\dot Q=0$.

\section{Higher-level Noether theorems}
\label{sec:higherlevel}

So far we have considered a general infinitesimal symmetry transformation, with constant parameter that we promote to an arbitrary function (subject to boundary conditions on its derivatives, when necessary) for the dual purpose of (i) reading off the corresponding Noether charge $Q$ and (ii) simpifying the proof that $Q$ is a constant of  motion.  We are now going to see
that a generalization of this procedure for higher-derivative theories leads to what could be called ``higher-level Noether theorems'', with corresponding ``Noetherian charges''. 

To this end, we consider a slightly relaxed version of  (\ref{highd})  so that the boundary conditions on derivatives of $\epsilon$ apply only for $\ell >1$: 
\begin{equation}\label{bc2}
0= \left. \frac{d^\ell \epsilon}{dt^\ell}\right|_{t=\alpha,\beta}\, \qquad  \ell =2, \dots, n-1\, .  
\end{equation}
Using this in (\ref{rewrite}) and setting $\Delta=0$ (since the possibility it represents will not be relevant to what follows) we have
\begin{equation}
\delta_\epsilon I  = \int_\alpha^\beta \left\{ \epsilon \left( h \, \frac{\delta I[x]}{\delta x(t)} + \dot Q_0\right) + \dot\epsilon Q_0\right\} dt +\left[ \dot\epsilon \bQ_1\right]_\alpha^\beta\, , 
\end{equation}
which we may rewrite as 
\begin{equation}
\delta_\epsilon I = \int_\alpha^\beta \left\{ \epsilon \left( h \, \frac{\delta I[x]}{\delta x(t)} + \dot Q_0\right) + \dot\epsilon \left(Q_0 + \dot \bQ_1\right) + \ddot\epsilon\,  \bQ_1 \right\} dt\, . 
\end{equation}

Now we suppose that $\delta_\epsilon I$  is zero not only for constant $\epsilon$ but also for constant $\dot\epsilon$.  The requirements for this are
\begin{equation}\label{extendedsym}
h \, \frac{\delta I[x]}{\delta x(t)} + \dot Q_0 =0\, , \qquad Q_0 + \dot \bQ_1 =0\, . 
\end{equation}
Given that these conditions are satisfied, we may promote $\epsilon$ to a function that is arbitrary except for the boundary conditions (\ref{bc2}), in which case
\begin{equation}
\delta_\epsilon I  = \int_\alpha^\beta \ddot\epsilon\,  \bQ_1 dt\, . 
\end{equation}
Proceeding by analogy with the simplified proof of Noether's theorem, we impose the additional boundary conditions
\begin{equation}
\epsilon(\alpha)=\epsilon(\beta) =0\, , \qquad \dot\epsilon(\alpha)=\dot\epsilon(\beta)=0\, ,
\end{equation}
and then integrate by parts to deduce that 
\begin{equation}
\delta_\epsilon I  = \int_\alpha^\beta \epsilon\,  \ddot{\bQ}_1  dt\, . 
\end{equation}
As the left hand side is zero for solutions of the EL equation, and $\epsilon(t)$ is constrained only at the integration limits, we deduce that 
\begin{equation}
\frac{\delta I[x]}{\delta x(t)}  =0 \quad \Rightarrow\quad \ddot{\bQ}_1=0 \, . 
\end{equation}
Notice that $\bQ_1$ is zero unless the action includes some higher-derivative term; more generally, $\bQ_p$ is zero unless $p< n$. 

Since $Q_0 + \dot \bQ_1=0$ is one of the conditions for invariance, an immediate corollary of $\ddot \bQ_1=0$ is $\dot Q_0$=0. In other words, $Q_0$
is a constant of motion; it is the Noether charge for the symmetry with $\dot\epsilon=0$, which is obviously a special case  of the symmetry with
$\ddot\epsilon=0$. Integrating $Q_0 + \dot \bQ_1=0$  we have $Q_0t +\bQ_1 = Q'$, another constant of motion. The two Noether charges
associated with invariance for constant $\dot\epsilon$ are therefore
\begin{equation}
Q_0 = -\dot\bQ_1 \qquad \&\qquad Q' = \bQ_1 + tQ_0\, . 
\end{equation}

A simple example  is provided by the higher order action
\begin{equation}\label{nice}
I[x] = \int_\alpha^\beta \tfrac{1}{2} \ddot x^2\, dt\, , 
\end{equation}
which is manifestly  invariant under the transformation $x\to x + \epsilon(t)$ for constant $\dot\epsilon$, i.e. when  $\ddot\epsilon=0$. The corresponding ``level-$1$''
Noetherian charge is 
\begin{equation}
\bQ_1 = \ddot x \, , 
\end{equation}
which indeed satisfies $\ddot\bQ_1=0$ as a consequence of the EL equation $\ddddot x=0$. The associated
two ``level-zero'' Noether charges are
\begin{equation}
Q_0= -\dot \bQ_1 = - \dddot x \qquad \& \qquad Q' = \bQ_1 + tQ_0 =  \ddot x -t \dddot x\, , 
\end{equation}
which are indeed both constants of the motion. Of course, these  can also be found by re-interpreting the transformation 
$x\to x + \epsilon(t)$ as the transformation $x\to x+ \epsilon\zeta(t)$ with {\it either} $\zeta=1$ {\it or} $\zeta= t$. 

There is an obvious generalization  to yet higher-level Noether theorems for which $\bQ_p$ 
is a  $p$th-level ``Noetherian charge''  satisfying $d^p\bQ_p/dt^p=0$ as a consequence of the EL equation. 
This implies that $\bQ_p$ is a $p$th-order polynomial in $t$ and the $p+1$ coefficients  of this polynomial are the Noether
charges required by Noether's theorem for a parameter $\epsilon(t)$ that is a $p$th-order polynomial in $t$.

The next step is to consider invariance of the action when  $\epsilon(t)$ is non-polynomial; then we have a {\it gauge invariance}
and Noether's ``second theorem''.

\section{Gauge invariance and higher derivatives}
\label{sec:gaugeinvariance}

We now turn to consider the implications, postponed from section \ref{sec:one}, of a zero Noether charge.  We shall continue to set $\Delta=0$, and we shall also replace the single function $x(t)$ by a set of 
functions $\{ x_1, x_2,\dots\}$ which we denote by ${\bf x}$.  Finally, for simplicity, we shall  restrict to the case of $n=2$; i.e. $L$ depends on ${\bf x}$ directly and through its first and second 
derivatives only.  In this case the variation $\delta_\epsilon I$ as given by the formula (\ref{DelI}) becomes
\begin{equation}\label{DelI2}
\delta_\epsilon I = \int_\alpha^\beta \left\{ \epsilon {\bf h} \cdot \frac{\delta I}{\delta {\bf x}}\right\} dt
+ \left[ \epsilon Q_0 + \dot\epsilon \bQ_1\right]_\alpha^\beta\, , 
\end{equation}
where
\begin{equation}
\frac{\delta I}{\delta {\bf x}} = \frac{\partial L}{\partial {\bf x}} - \frac{d}{dt} \left(\frac{\partial L}{\partial \dot{\bf x}}\right) + \frac{d^2}{dt^2} \left(\frac{\partial L}{\partial\ddot {\bf x}}\right)\, , 
\end{equation}
and 
\begin{eqnarray}
Q_0 &=& \xi L + {\bf h}\cdot \left[\frac{\partial L}{\partial \dot{\bf x}} - \frac{d}{dt} \left(\frac{\partial L}{\partial \ddot{\bf x}}\right)\right] + \dot{\bf h} \cdot 
\frac{\partial L}{\partial \ddot{\bf x}}\, , \\
\bQ_1 &=& {\bf h}\cdot \frac{\partial L}{\partial \dot{\bf x}}\, . 
\end{eqnarray}
The dot product notation here is shorthand for a sum over the components of ${\bf x}$, and we remind the reader that
\begin{equation}
{\bf h} = \bfz- \xi\dot{\bf x}\, , 
\end{equation}
where $\xi$ and $\bfz$ are the functions that determine the infinitesimal transformation (\ref{transform}).

Let us consider first the special case in which  $L$ has no dependence on $\ddot{\bf x}$; i.e. no higher derivatives. In this case we may rewrite
(\ref{DelI2}) as 
\begin{equation}\label{DelI3}
\delta_\epsilon I = \int_\alpha^\beta \left\{ \epsilon \left({\bf h} \cdot \frac{\delta I}{\delta {\bf x}} + \dot Q_0\right) + \dot\epsilon Q_0 \right\} dt\, . 
\end{equation}
For $\dot\epsilon=0$ we recover the condition (\ref{symcon}) for invariance of the action in the form 
\begin{equation}\label{invariance}
{\bf h} \cdot \frac{\delta I}{\delta {\bf x}} + \dot Q_0 =0\, . 
\end{equation}
Allowing for non-constant $\epsilon$ we read off the Noether charge $Q_0$ from the coefficient of $\dot\epsilon$. 
However, if $Q_0=0$ then (\ref{invariance}) becomes the condition for invariance of $I$  for {\it arbitrary} $\epsilon(t)$; i.e.  a gauge transformation. 
Moreover, this condition reduces for $Q_0=0$ to the equation
 \begin{equation}\label{Bid}
{\bf h} \cdot \frac{\delta I}{\delta {\bf x}} =0\, , \end{equation}
which implies that not all components of the EL equation $\delta I/\delta{\bf x}={\bf 0}$ are linearly independent; i.e. they are subject to a ``Noether identity''. 
This is Noether's ``second theorem''.

Now we aim to consider the implications of higher derivative terms, so we reinstate the dependence of $L$ on $\ddot{\bf x}$, and rewrite 
(\ref{DelI2}) as 
\begin{equation}\label{n=2}
\delta_\epsilon I = \int_\alpha^\beta \left\{ \epsilon \left({\bf h} \cdot \frac{\delta I}{\delta {\bf x}} + \dot Q_0\right) + 
\dot\epsilon Q_0 \right\} dt  + \left[\dot\epsilon\, \bQ_1\right]_\alpha^\beta\, . 
\end{equation}
This differs from (\ref{DelI3}) only by the addition of a boundary term on the right hand side, so the implications of $Q_0=0$ are unchanged from the discussion above if we  restrict 
$\epsilon(t)$ to satisfy
\begin{equation}\label{restrict1}
\dot\epsilon(\alpha)=\dot\epsilon(\beta) =0\, . 
\end{equation}
As this imposes only two conditions on an otherwise arbitrary function, we still have a gauge invariance, albeit a restricted one. 

If the condition  (\ref{restrict1}) is not imposed then we can rewrite (\ref{n=2}) as 
\begin{equation}
\delta_\epsilon I = \int_\alpha^\beta \left\{ \epsilon \left({\bf h} \cdot \frac{\delta I}{\delta {\bf x}} + \dot Q_0\right) + 
\dot\epsilon \left( Q_0 + \dot\bQ_1\right) + \ddot\epsilon \bQ_1 \right\} dt \, . 
\end{equation}
The condition for invariance when $\epsilon$ is constant is again (\ref{invariance}). The additional condition required for invariance when $\dot\epsilon$ is 
constant is $Q_0 + \dot\bQ_1=0$; this is the case discussed in the previous section. Finally, we have an {\it unrestricted} gauge invariance when 
both $Q_0$ and $\bQ_1$ are zero.  More generally, for $n>2$, unrestricted gauge invariance requires 
\begin{equation}
Q_0= 0\, \qquad \& \qquad \qquad \bQ_p=0\, , \quad p=1, \cdots, n-1\, . 
\end{equation}

Consider, for example, the following time reparametrisation invariant action relevant to the dynamics of a point particle  \cite{Pisarski:1986gp}
\begin{equation}
I [{\bf x}] = \int \frac{\sqrt{(\dot{\bf x}\cdot\ddot{\bf x})^2 - \dot{\bf x}^2 \ddot {\bf x}^2}}{\dot{\bf x}^2}\, dt\, . 
\end{equation}
This  is invariant for arbitrary $\epsilon(t)$ under an infinitesimal transformation with
$(\xi,\bfz)=(1,{\bf 0})$ (and hence ${\bf h} = \dot{\bf x}$). Using the identities
\begin{equation}
\dot{\bf x} \cdot \frac{\partial L}{\partial \ddot {\bf x}} \equiv 0\, , \qquad  \ddot{\bf x} \cdot \frac{\partial L}{\partial \dot {\bf x}} \equiv 0\, , 
\end{equation}
and 
\begin{equation}
\dot{\bf x} \cdot \frac{\partial L}{\partial \dot {\bf x}} + L \equiv 0\, , \qquad 
\ddot{\bf x} \cdot \frac{\partial L}{\partial \ddot {\bf x}} - L \equiv 0\, , 
\end{equation}
one may verify that $Q_0=\bQ_1=0$.

\section{Discussion}

A proof of Noether's theorem has been a standard item  in  Quantum Field Theory texts for a long time but the standard proof of it  has been displaced within the last 25 years by a simpler proof that 
appears to have its origins in the work of Gell-Mann and Levy in 1960 \cite{GellMann:1960np}. This simpler proof starts by ``promoting'' the constant parameter of a symmetry transformation to a 
function of time (spacetime in the field theory context) and a significant practical implication  is that the Noether charge (current in the field theory context) can be read off from the
time derivative (spacetime derivative for field theory) of the promoted parameter in the variation of the action, thus allowing an expression for the Noether charge to be extracted from 
a calculation designed to verify  invariance. 

This idea was widely used in the construction of supersymmetric field theories, including supergravity theories, starting in the mid-1970s, when the 
simpler proof of Noether's theorem became well-known in that community. However, there does not appear to have been any attempt  to put this simpler proof 
on a firmer foundation,   e.g. by a detailed analysis of boundary terms in the variation of the action or by consideration of its applicability to higher-derivative actions. 
This paper has addressed both these issues,  admittedly  in the context of Lagrangian  mechanics but field theory is (at least in principle) just mechanics for a system with a continuous infinity of variables. 

The initial  motivation for this paper was a suspicion that the simplified proof of Noether's theorem would require modification for higher-derivative actions. 
The theorem itself was never in doubt because Noether allowed for higher derivatives in her original proof. However,   the simpler derivation of it for higher-derivative actions
does require additional boundary conditions on the ``promoted'' parameter;  without them the varied action  has {\it either} boundary terms  assumed absent in the simplified proof 
{\it or}  terms  proportional to $\ddot\epsilon$, and possibly higher time derivatives of $\epsilon$, in addition to the term proportional to $\dot\epsilon$.

The details of the simplified proof of Noether's theorem are therefore slightly different for higher-derivative actions. The difference could be viewed as a technicality,
at least for actions that involve no higher than $n$th derivatives for some {\it finite} $n$,  because it is just  a matter of having to impose boundary conditions on a finite number
of derivatives of the promoted parameter $\epsilon(t)$. However, for an effective action that is an infinite sum of terms of increasingly higher-order in derivatives, it might be necessary 
to impose boundary conditions on {\it all} derivatives of $\epsilon(t)$, and the implications of this are less clear. 

The differences introduced by higher-derivatives also suggest a generalisation of Noether's theorem. 
For a higher-derivative action it may happen that a transformation with parameter $\epsilon$ is a symmetry not only when $\epsilon$ is constant but also when $\dot\epsilon$ 
is constant. In this case, the terms in the varied action proportional to $\epsilon$ {\it and} $\dot\epsilon$ will be zero but there will still be a term proportional to  
$\ddot\epsilon$, with coefficient $\bQ_1$.   A close analog  of the simplified proof of Noether's theorem can now be  used to show that $\ddot\bQ_1=0$ as a consequence of the equations 
of motion; we called $\bQ_1$ a ``level-one  Noetherian charge''. More generally, for an action that is invariant under a symmetry transformation with parameter $\epsilon(t)$ that is a $p$th order polynomial in $t$, there is a ``level-$p$ Noetherian charge'' that is, as a consequence of the equations of motion, a $p$th-order polynomial in $t$ with coefficients that are all standard Noether charges\footnote{It is possible that this
idea is related to  the jet-bundle approach to variational principles for higher-derivative theories \cite{Aldaya:1978uj}.}.

What is the field theory analog of  ``higher-level Noetherian charges''? Consider the field theory analog of the relation $Q_0+ \dot\bQ_1=0$ of (\ref{extendedsym}): as $Q_0$ is replaced by a
Noether current $J^\mu$ ($\mu=0,1, \dots$) and the time derivative by a spacetime derivative, we must replace $\bQ_1$ by a symmetric tensor $\bJ_1^{\mu\nu}$ such that 
\begin{equation}\label{firstint}
J^\mu + \partial_\nu \bJ_1^{\mu\nu} =0\, . 
\end{equation}
The conservation condition $\partial_\mu J^\mu=0$ now implies that $\partial_\mu\partial_\nu \bJ^{\mu\nu}=0$, which is the field theory analog of $\ddot\bQ_1=0$. The first integral of this equation is 
(\ref{firstint}). Integrating again we have
\begin{equation}
\bJ_1^{\mu\nu} = T^{\mu\nu}  - 2 x^{(\mu} J^{\nu)} \, , \qquad \partial_\mu T^{\mu\nu} =0\, . 
\end{equation}
An obvious candidate for the symmetric tensor $T^{\mu\nu}$ is the stress-energy tensor. With this interpretation, the Noetherian tensor $\bJ^{\mu\nu}$ unifies the stress-energy tensor with an abelian 
Noether current, e.g. electric charge current. The extension to level-2 and beyond is probably less interesting since it would require at least a spin-3 symmetry that is only possible for a free field theory. 
There may be a relation between these ideas  and the jet-bundle approach to variational principles for higher-derivative theories \cite{Aldaya:1978uj}

Returning to mechanics, an invariance of the action for {\it arbitrary} $\epsilon(t)$ is a gauge invariance, and a gauge invariance is not a symmetry because the Noether charge 
$Q$ is zero.  There is a modification to this standard statement in the presence of higher-derivatives because then $Q=0$ implies a gauge invariance only if derivatives of $\epsilon$ are subject
to boundary conditions; we have called this  a ``restricted gauge invariance''.   For unrestricted gauge invariance, which we argued to be the more physically relevant possibility, 
at least for mechanics,  all higher-level Noetherian charges must be zero too. The field theory analog of an unrestricted gauge invariance would require, for a higher-derivative theory, at least 
a spin-3 gauge invariance.  

The issues raised by higher-derivative actions that have been discussed in this paper could have been addressed in a different way.  We could have first 
converted a given higher-derivative action to an equivalent action without higher-derivatives by the introduction of additional ``auxiliary'' variables. Then, it would appear, 
we could apply the simplified proof of Noether's theorem without having to concern ourselves with boundary conditions on derivatives of $\epsilon$, and all higher-level
Noetherian charges would be identically zero. We conclude with an examination of  this point in the context of the simple  higher-derivative action of (\ref{nice}). The variation 
of this action for transformation $x\to x+ \epsilon(t)$ is 
\begin{equation}\label{discuss}
\delta_\epsilon I[x] = \int \dot\epsilon Q_0 + \left[\dot\epsilon\bQ_1\right]\, , 
\end{equation}
where $Q_0$ is  the Noether charge associated to invariance for $\epsilon=\epsilon_0$, a constant.  As explained in section \ref{sec:higherlevel}, $\bQ_1$ is a ``level-one Noetherian charge'' that is simply related to a second Noether charge $Q'$ associated to invariance for $\epsilon = \epsilon_1t$ for constant $\epsilon_1$.  However, as explained in section \ref{sec:one}, the simplified proof of 
Noether's theorem will apply to the symmetry $x\to x+ \epsilon_0$ only if we first remove the boundary term in $\delta_\epsilon I[x]$ by setting  $\dot\epsilon$ to zero at the integration limits.
The question we want to answer is whether this restriction on $\epsilon$ can be evaded by the introduction of auxiliary variables.

Consider first the following action that is manifestly equivalent to $I[x]$ of (\ref{nice}): 
\begin{equation}
I[x,y] = \frac{1}{2} \int \left\{ \ddot x^2 - \left(y- \ddot x\right)^2\right\} dt \ = \,  -\int \left\{\dot x\dot y + \frac{1}{2} y^2\right\}dt \ +\, \left[\dot x y\right]\, .  
\end{equation}
The higher-derivative terms have cancelled, after an integration by parts, but this step has introduced a boundary term in the expression for $I[x,y]$. Taking this boundary contribution into account, we find that the variation of $I[x,y]$  for transformation $x\to x+ \epsilon(t)$ is 
\begin{equation}
\delta_\epsilon I[x,y] = - \int \dot\epsilon \dot y\, dt + \left[ \dot\epsilon y\right] = \delta_\epsilon I[x]\, , 
\end{equation}
where we use $y= \ddot x$ (the EL equation for $y$) to arrive at the second equality.  Not surprisingly, an equivalent action has yielded equivalent results. 

Undeterred, we might now  consider omitting the boundary term $[\dot xy]$ in  $I[x,y]$ to arrive at the new action\footnote{The new action will require new boundary conditions on $(x,y)$ for 
functional derivatives to be defined, but we pass over this difference here.}
\begin{equation}
\tilde I[x,y] = -\int \left\{\dot x\dot y + \frac{1}{2} y^2\right\}dt \, . 
\end{equation}
For the transformation $x\to x+ \epsilon(t)$ we now have 
\begin{equation}
\delta_\epsilon \tilde I =  -\int \dot \epsilon \dot y dt  = \int \dot\epsilon Q_0\, , 
\end{equation}
where we again use $y= \ddot x$ for the second equality. The boundary term $[\dot\epsilon\bQ_1]$ has gone, so the simplified proof of  Noether's theorem now applies without any restriction on $\dot\epsilon$, and it  tells us that $Q$ is the Noether charge corresponding to  the symmetry $x\to x+\epsilon$ for constant $\epsilon$. However, for  constant $\dot\epsilon$ we now have 
$\delta_\epsilon \tilde I = -\left[ \dot\epsilon y\right]$, so that $\tilde I[x,y]$ is not invariant (in contrast to $I[x,y]$ because the two terms of (\ref{discuss}) cancel when  
$\ddot\epsilon=0$). In fact, setting $\epsilon(t) = vt$ for constant parameter $v$, we  have
\begin{equation}
\delta_v \tilde I =  -\left[vy\right] \qquad \Rightarrow \quad \Delta = -vy\, . 
\end{equation}
Applying the formula (\ref{Q}), which allows for $\Delta\ne0$,  yields the Noether charge $Q'$. 

On the basis of this example, it appears plausible that a higher-derivative action with a symmetry for $p$th-order polynomial parameter $\epsilon(t)$ could be replaced by an action depending on more variables 
but without higher-derivatives, and that in this case the extra $p$  Noether charges  will  correspond to symmetries of the equations of motion that are not symmetries of the action.

\section*{Acknowledgements}
The author is grateful to Alex Arvanitakis for helpful discussions, and to Jos\'e de Azc\'arraga, Bernard de Wit and Peter van Nieuwenhuizen for helpful correspondence. 
Support from the UK Science and Technology Facilities Council (grant ST/P000681/1) is gratefully acknowledged.

\providecommand{\href}[2]{#2}\begingroup\raggedright\endgroup


\end{document}